\pdfoutput=1 % Uncommented for upload to arXiv.org
\documentclass[applsci,article,submit,pdftex,moreauthors]{Definitions/mdpi} 

\usepackage{epstopdf}
\usepackage{amsmath,amssymb,amsfonts}
\usepackage{graphicx}
\makeatletter
\let\c@lofdepth\relax
\let\c@lotdepth\relax
\makeatother
\usepackage{subfigure}
\usepackage[T1]{fontenc}
\usepackage{textcomp}
\usepackage{xcolor}
\usepackage{algcompatible}
\usepackage{algorithm}

\def\BibTeX{{\rm B\kern-.05em{\sc i\kern-.025em b}\kern-.08em
    T\kern-.1667em\lower.7ex\hbox{E}\kern-.125emX}}
\firstpage{1} 
\pubvolume{1}
\issuenum{1}
\articlenumber{0}
\pubyear{2023}
\copyrightyear{2023}
\datereceived{ } 
\daterevised{ } % Comment out if no revised date
\dateaccepted{ } 
\datepublished{ } 
\hreflink{https://doi.org/} % If needed use \linebreak
\Title{CRC-based Reliable WiFi Backscatter Communiation for Supply Chain Management}

\TitleCitation{Title}

\Author{Yun-Hao Liu $^{1}$, Tao Liu $^{2}$, Yimeng Huang $^{1}$, Han Ding $^{3}$, Wei Xi $^{3}$ and Wei Gong $^{1,}$*}

\AuthorNames{Yun-Hao Liu, Tao Liu, Yimeng Huang, Han Ding,  Wei Xi and Wei Gong}

\AuthorCitation{Liu, Y.; Liu, T.; Huang, Y.; Ding H.; Xi W.; Gong W.}
\address{%
$^{1}$ \quad School of Computer Science and Technology, University of Science and Technology of China, Hefei, 230026, An
hui, China; zkdjy2018lyh@mail.ustc.edu.cn(Y.L.); huangyimeng@mail.ustc.edu.cn(Y.H.); weigong@ustc.edu.cn(W.G.)\\
$^{2}$ \quad Business School, China University of Political Science and Law, 102249, Beijing, China; Liuliutao@cutech.edu.cn(T.L.)\\
$^{3}$ \quad School of Computer Science and Technology, Xi'an Jiaotong University, Xi'an, 710049, Shaanxi, China; dinghan@xjtu.edu.cn(H.D.); xiwei@xjtu.edu.cn(W.X.)}

\corres{Correspondence: weigong@ustc.edu.cn}

\preto{\abstractkeywords}{\nolinenumbers}
\abstract{Supply chain management is aimed to keep going long-term performance of the supply chain and minimize the costs. Backscatter technology provides a more efficient way of being able to identify items and real-time monitoring.  Among the backscatter systems, the ambient backscatter communication (AmBC) system provides a prospect of ultra-low energy consumption and does not require controlled excitation devices. In this paper, we introduce CRCScatter, a CRC reverse algorithm-based AmBC system using a single access point (AP). A CRC reverse decoder is applied to reverse the ambient data from CRC32 sequence in the backscatter packet and realize single-AP decoding. Based on the nature of DBPSK modulation in WiFi signal, the CRCScatter system obtains the tag data by XOR and Differential decoder.
Our simulation results verify the effectiveness of our proposed system in the low SNR regime. The average decoding time of CRCScatter system is independent of the length of tag data. 
Furthermore, our system can append redundant bits in the tag data to improve the decoding accuracy while not increasing the decoding time.}


\begin{document}

%%%%%%%%%%%%%%%%%%%%%%%%%%%%%%%%%%%%%%%%%%
% \setcounter{section}{-1} %% Remove this when starting to work on the template.
% \section{How to Use this Template}

% The template details the sections that can be used in a manuscript. Note that the order and names of article sections may differ from the requirements of the journal (e.g., the positioning of the Materials and Methods section). Please check the instructions on the authors' page of the journal to verify the correct order and names. For any questions, please contact the editorial office of the journal or support@mdpi.com. For LaTeX-related questions please contact latex@mdpi.com.%\endnote{This is an endnote.} % To use endnotes, please un-comment \printendnotes below (before References). Only journal Laws uses \footnote.

% The order of the section titles is different for some journals. Please refer to the "Instructions for Authors” on the journal homepage.

\section{Introduction}

In the next generation Internet of Things, billions of devices will connect to existing networks for data communications. 
In the highly competitive industry services, supply chain management has very high requirements for informatization.
The monitoring and communication equipment in the supply chain will make power consumption a serious challenge.
Enterprises that give priority to the application of low energy consumption equipment and technology in the supply chain will gain higher profits.
The power consumption of these devices will become a critical problem. According to a survey of wireless communication \cite{b1}, most of the energy consumed by active communication systems is spent on energy-hungry active RF devices such as power amplifiers. 

To reduce energy consumption and provide long-time communication in passive conditions, a novel low-cost and energy-efficient communication system, backscatter communication system was proposed and could reduce power consumption to the microwatt level. These advantages make backscattering promising for large-scale applications in smart home \cite{b2}, smart agriculture \cite{b3}, and other fields. So far, backscatter has been implemented in varieties of radios, including WiFi \cite{b4,b5,b6}, ZigBee \cite{b7}, LoRa \cite{b8} and Bluetooth \cite{b9,b10,b11}.
Traditional backscatter systems require specialized hardware to achieve backscatter communication. 
WiFi backscatter \cite{b12} requires a power supply to support the system to connect to the network. BackFi \cite{b13} applies customized full duplex hardware to enable backscatter communication. Passive WiFi \cite{b14} requires a dedicated continuous wave signal generator as the excitation signal source. These additional requirements on hardware will limit the application scenarios of the backscatter system. HitchHike \cite{b15} achieves backscatter communication by commodity devices and introduces the idea of “codeword translation” which allows backscatter tag to embed its information on standard 802.11b packets. Codeword translation then has been extended to Bluetooth, ZigBee, and LoRa by FreeRider \cite{b16} and LoRa Backscatter \cite{b17}. In detail, HitchHike system deploys one more access point to demodulate the excitation signal and the backscatter signal respectively, which enables the system to decode the tag data by a simple and efficient XOR decoder.

% \begin{figure}[H]
% \includegraphics[width=10.5 cm]{Definitions/logo-mdpi}
% \caption{This is a figure. Schemes follow the same formatting. If there are multiple panels, they should be listed as: (\textbf{a}) Description of what is contained in the first panel. (\textbf{b}) Description of what is contained in the second panel. Figures should be placed in the main text near to the first time they are cited. A caption on a single line should be centered.\label{fig1}}
% \end{figure} 

\begin{figure}[H]
\subfigure[backscatter system model A]{
\includegraphics[width=8cm]{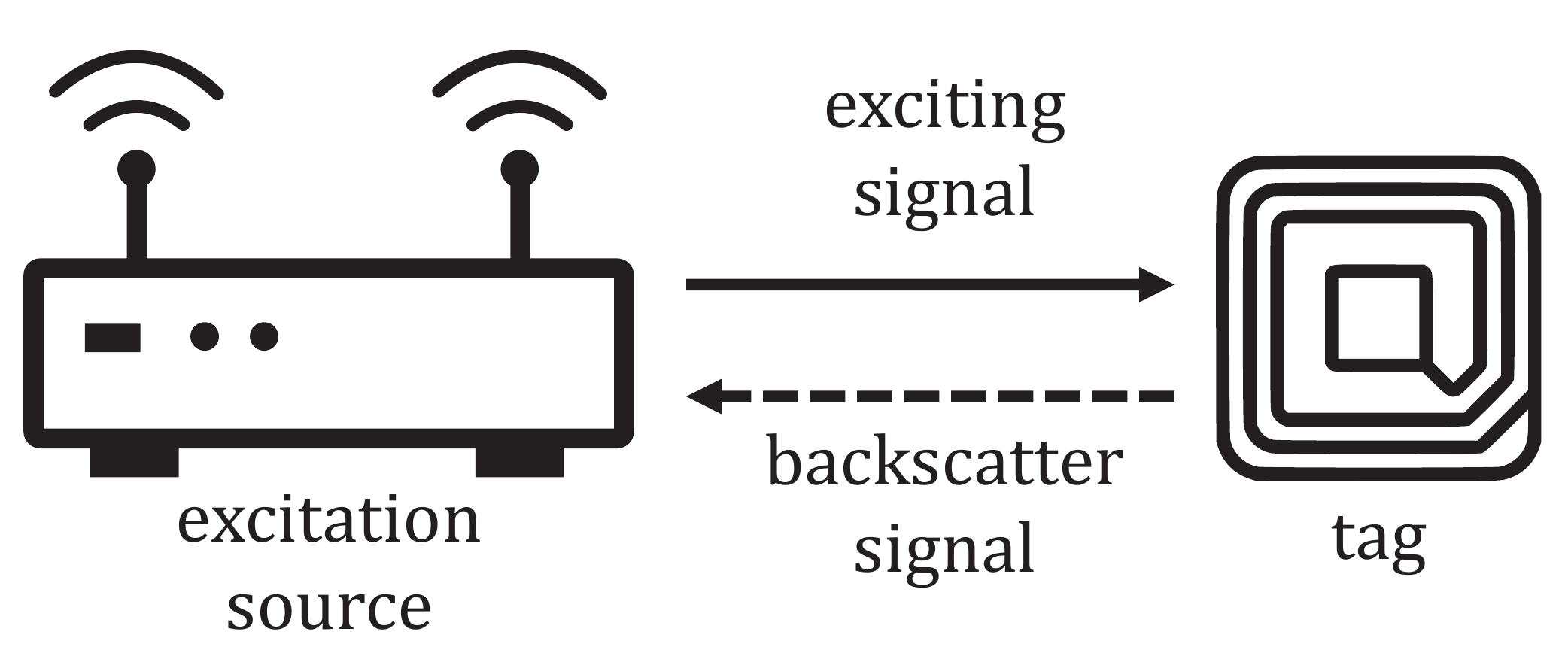}
}

\subfigure[backscatter system model B]{
\includegraphics[width=12cm]{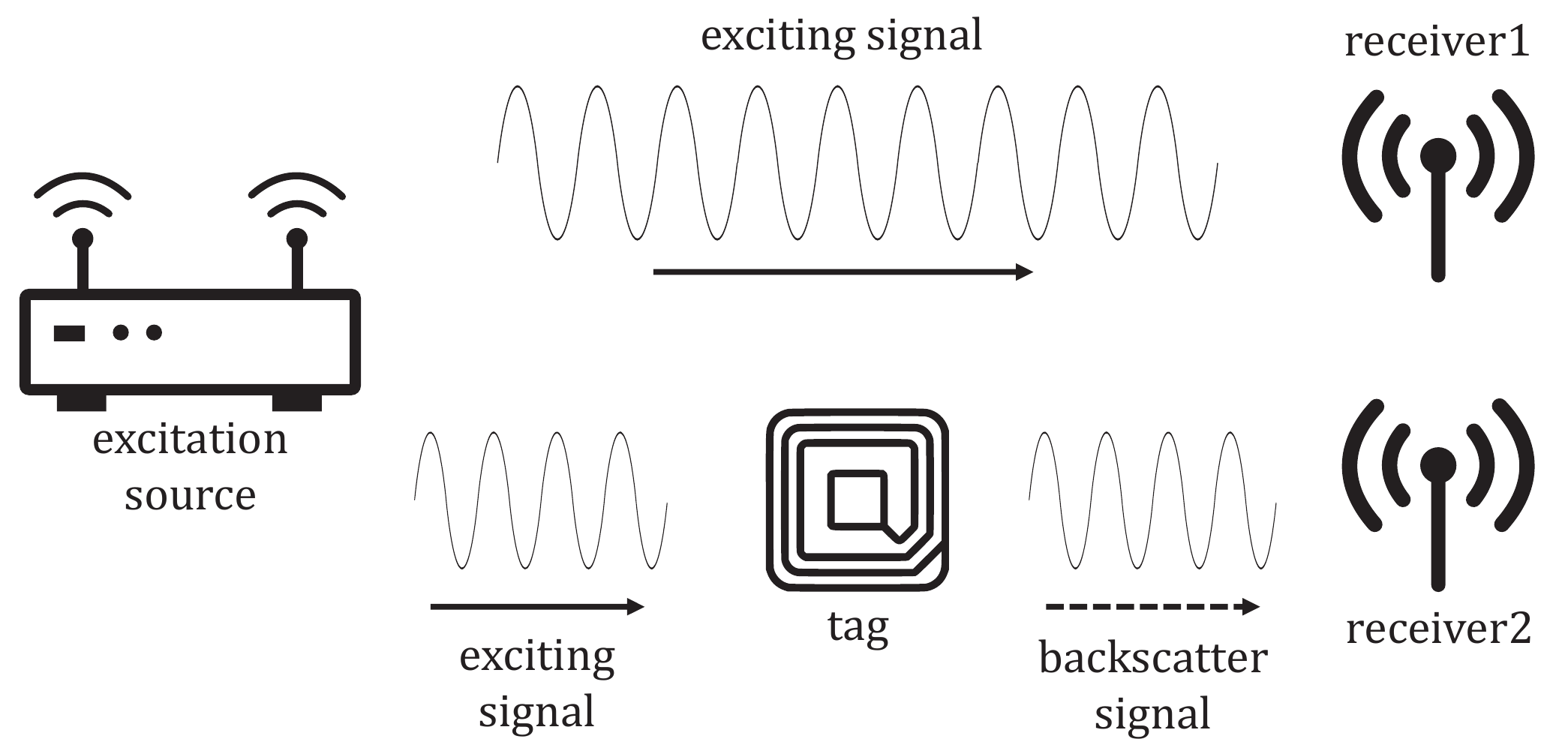}
}
\caption{The structure of backscatter communication systems.\label{fig1}}
\end{figure}

We can summarize the backscatter systems as two modes in Figure~\ref{fig1}. 
In model A, the backscatter communication system with a controlled excitation source enables decoding at a single-AP receiver while employing additional requirements on the excitation source. The controlled excitation source signal does not contain valid information but adds interference to other communications. 
In model B, the backscatter communication system deploys one more access point (AP) to receive the excitation signal to decode tag data directly using the backscatter data and original data. 
In general, compared with the active communication infrastructure, these two backscatter system models have additional hardware requirements at the transmitter or receiver. 
The deployment of systems will cause extra costs and face difficulties in the large-scale deployment on current WiFi infrastructure.

Inspired by these observations, this paper focuses on single-AP decoding with ambient WiFi signals. The main contributions in this work are summarized as follows.
\begin{itemize}
\item We propose a CRC reverse algorithm-based single-AP backscatter system using ambient WiFi signal.  The system uses a CRC reverse decoder to solve the problem of decoding ambient data from a backscatter packet.
\item We provide simulation results to verify the decoding method of our system and analyze the system performance. The CRCScatter system achieves decoding tag data bit error rate of $10^{-2}$ at SNR = $-7.5$ dB.
\item We verify that the CRC reverse algorithm is better than the brute-force search method in decoding efficiency and the average decoding time of CRCScatter system is independent of the tag data length. We propose an improved method of adding redundant bits to the tag data to improve the decoding accuracy of the system in the presence of noise interference.
\end{itemize}

%%%%%%%%%%%%%%%%%%%%%%%%%%%%%%%%%%%%%%%%%%

\section{System Model}

\begin{figure}[H]
\includegraphics[width=13 cm]{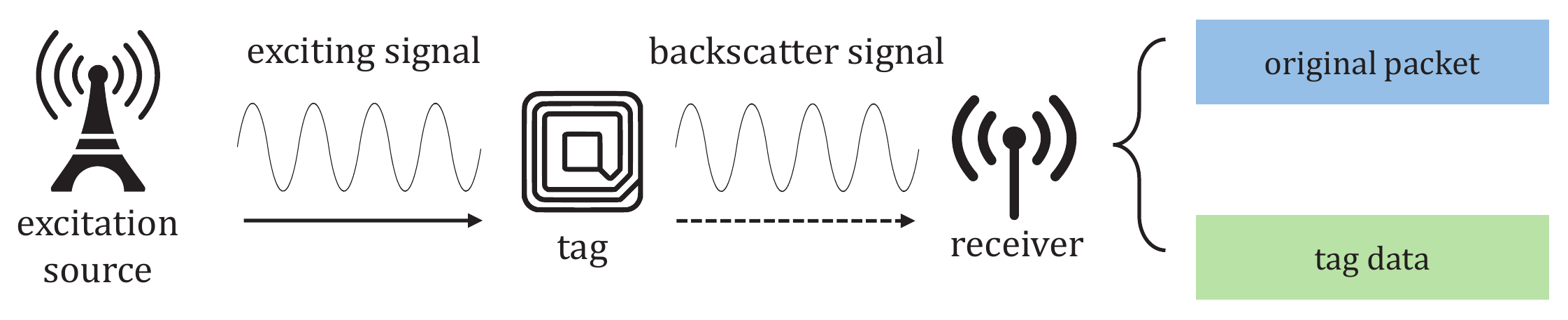}
\caption{CRCScatter system model.\label{fig2}}
\end{figure}

The system model for the CRCScatter system consists of an excitation source, a tag, and a single-AP receiver as shown in Figure~\ref{fig2}. 
The ambient excitation source broadcasts the exciting signal to the tag of the CRCScatter system. 
The tag transmits its data by backscattering and modulating the exciting signal. 
The CRCScatter receiver uses only one access point to receive the backscatter signal from the tag to the receiver.
Considering the tag data transmission, the tag data is encoded on the WiFi signal by tag modulation and the receiver uses the backscatter packet to reverse the original packet and decode the tag data.
In this section, we present how the tag piggybacks data on a backscatter signal using codeword translation.
This paper focuses on the 1 Mbit/s data rate and other data rates will be implemented in our future work.

\subsection{802.11b packet structure}
\begin{figure}[H]
\includegraphics[width=12 cm]{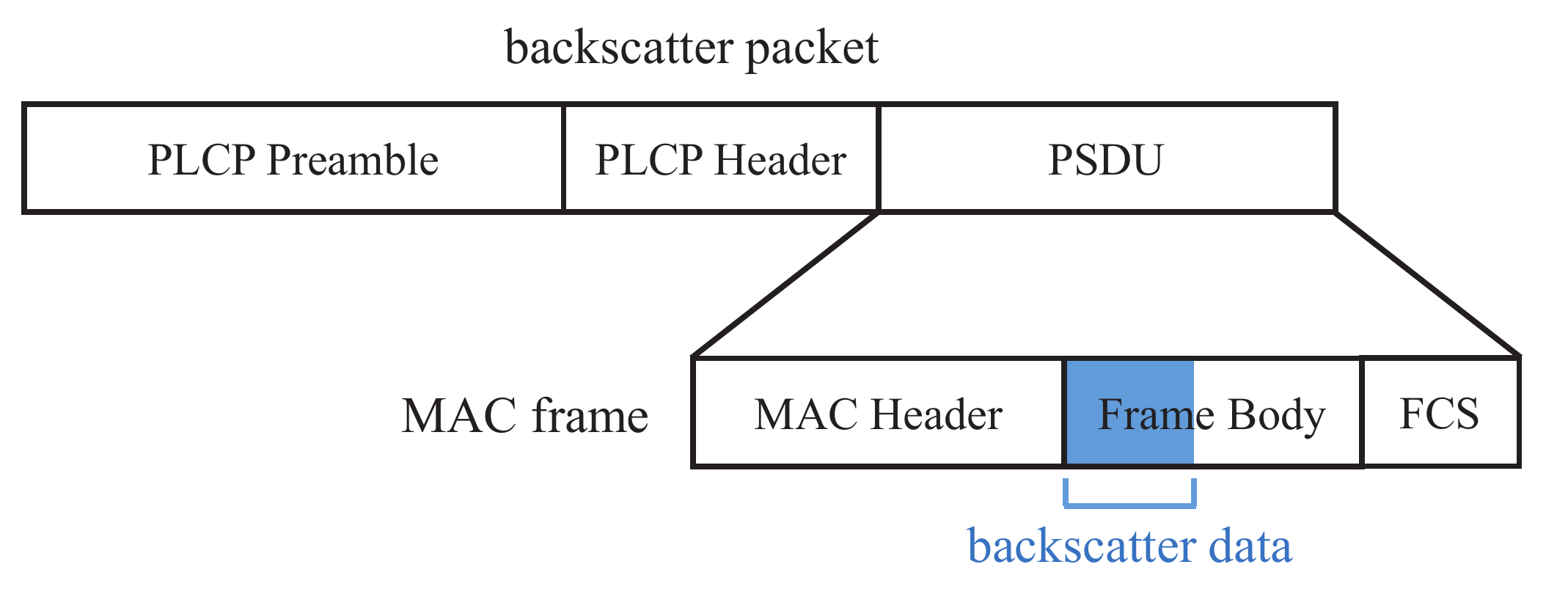}
\caption{The structure of received backscatter packet.\label{fig3}}
\end{figure}

The 802.11b protocol specifies that the transmitted packet contains PLCP Preamble, PLCP Header, and PSDU. 
Among them, PSDU has variable length and contains a MAC frame in the CRCScatter system.
The MAC frame consists of MAC Header, Frame Body, and the frame check sequence (FCS). 
The Frame Body field carries the transmitted data and the CRC32 sequence in FCS field can detect bit errors in the MAC frame.
The part of the packet other than the data segment contains packet control information, if these fields are modified by tag, then the receiver may not be able to demodulate correctly.
To receive and demodulate properly, the CRCScatter tag only modulates the Frame Body field in the packet. 
The received backscatter packet structure is shown in Figure~\ref{fig3}.

\subsection{Codeword translation}

A novel technology called codeword translation is proposed in HitchHike \cite{b15} for tag modulation. 
The codeword translation method enables the piggybacking of tag data information while ensuring that the backscatter packet is still an 802.11b WiFi packet that can be demodulated by the receiver.
Specifically, the 802.11b 1Mbps signal uses two codewords to encode packets and there is a 180° phase difference between these two codewords.
The tag adopts BPSK modulation to change the codeword by a phase offset during modulating and backscattering the exciting signal exhibited in Table~\ref{tab1}. 
\begin{table}[H] 
\caption{Encoding at the tag.\label{tab1}}
\newcolumntype{C}{>{\centering\arraybackslash}X}
\begin{tabularx}{\textwidth}{CCC}
\toprule
\textbf{tag bit}	& \textbf{phase offset}\\
\midrule
0		& 0\\
1		& 180°\\
\bottomrule
\end{tabularx}
\end{table}

In tag modulation, one tag data bit corresponds to one data bit.
This correspondence indicates that the encoding scheme is efficient and redundancy-free.
HitchHike system proposed an efficient XOR decoder to get the tag data from the backscatter data and original data. 
The decoding formula of the XOR decoder can be written as:
\begin{linenomath}
\begin{equation}
tag\ data=backscatter\ data \oplus original\ data.
\end{equation}
\end{linenomath}

However, in the CRCScatter system there is only one AP for receiving signals. 
The backscatter data for the XOR decoder can be obtained directly, while the original data need to be calculated from the backscatter packet by the CRCScatter decoder. 
Thus the decoding function of the CRCScatter system can be expressed as follows:

\begin{linenomath}
\begin{equation}
(original\ packet,tag\ data)=CRCScatter(backscatter\ packet).
\end{equation}
\end{linenomath}

\section{CRCScatter decoder design}

\subsection{CRCScatter decoder overview}

\begin{figure}[H]
\includegraphics[width=12 cm]{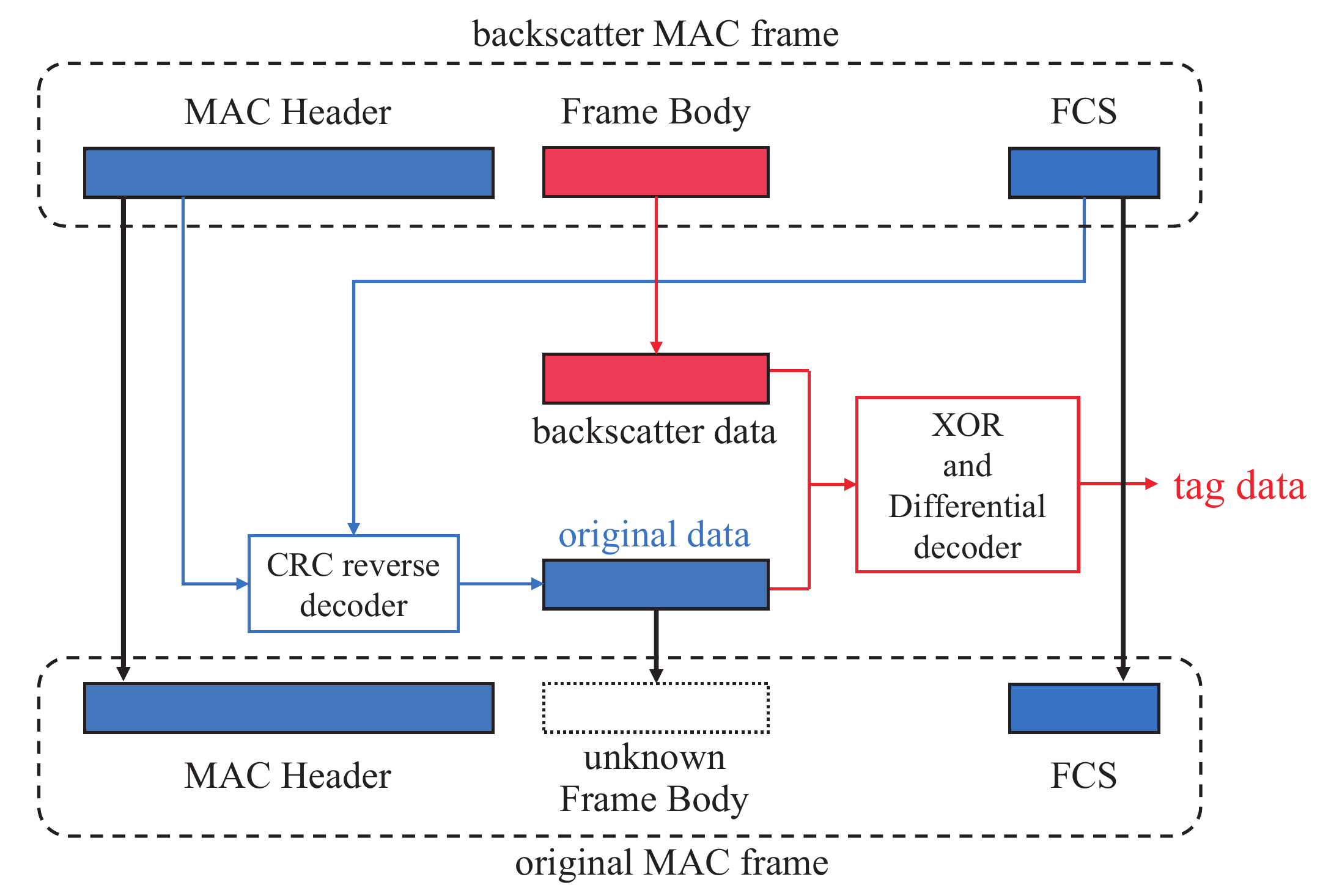}
\caption{CRCScatter decoder overview. The decoding procedure can be divided into two steps. The first step is to reverse the original data marked in blue, and the second step is to calculate the tag data marked in red.\label{fig4}}
\end{figure} 

The overview of the CRCScatter's decoding method at the receiver is illustrated in Figure~\ref{fig4}. 
The procedure of tag data decoding by the CRCScatter decoder can be divided into two steps.

The first step is to reverse the backscatter packet into the original packet.
The received backscatter MAC frame contains MAC Header, Frame Body, and FCS, but the tag only modulates the Frame Body field. 
The MAC Header field and FCS field in the original packet are the same as those in the backscatter packet.
Nevertheless, the original data cannot be obtained directly because the Frame Body field has been modulated by the tag.
The 802.11b protocol specifies the correlation between the FCS field and transmitted data in the 802.11b packet.
CRC reverse decoder uses this intrinsic correlation of the 802.11b packet as a constraint in calculation and utilizes two CRC algorithms to reverse the original data from the backscatter packet.

The second step is to calculate the tag data based on the received backscatter data and the original data obtained by the CRC reverse decoder.
The 802.11b data transmission uses DBPSK modulation, while the tag adopts BPSK modulation.
This difference in modulation prevents the original XOR decoder from getting the correct tag data.
Therefore, CRCScatter uses a decoder that combines XOR operation and differential decoding to calculate the tag data.

In summary, the CRCScatter system deploys two decoders to calculate the original packet and tag data in steps.
The function of the CRC reverse decoder is to get the original data from the backscatter packet, 
and the XOR and Differential decoder removes the differential effects in modulation to obtain the correct tag data.

\subsection{CRC reverse decoder}
\subsubsection{The algorithms of CRC in the CRC reverse decoder}

To detect the unpredictable bit errors in the received packet, Cyclic Redundancy Check (CRC) sequence is invented and transmitted with the data.
In the 802.11b packet, the CRC32 is applied in the FCS field of the MAC frame to protect the MAC Header and Frame Body fields.
The CRC32 value can be calculated by bit shift and XOR operation on a 32-bit CRC register.
The bit-oriented calculation algorithm of the CRC value is given by the 802.11b protocol shown in Algorithm 1.
The CRC algorithm contains three constants: CRCPOLY, INITXOR, and FINALXOR.
The value of the constants in CRC32 is given by:
CRCPOLY = 0x04C11DB7, INITXOR = FINALXOR = 0xFFFFFFFF.
Moreover, we can use the initial and final state of the $crcreg$ to replace INITXOR and FINALXOR when focusing on the calculation of the CRC register.

\begin{algorithm}
\caption{calculation of the CRC}
\begin{algorithmic}
\REQUIRE{data bits $a$}
\ENSURE{CRC register value $crcreg$}
 \STATE $crcreg \gets $ INITXOR
 \STATE $i \gets 0$
 \WHILE{$i\ <\ a.length$}
  \STATE LEFTSHIFT($crcreg$)
  \IF{$bit\_just\_shifted\_out \neq a_{i}$}
   \STATE $crcreg \gets {crcreg\ \oplus  }$\ CRCPOLY 
  \ENDIF
  \STATE $i \gets i+1$
 \ENDWHILE
 \STATE $crcreg \gets ${$crcreg\ \oplus\ $FINALXOR} 
\end{algorithmic}
\end{algorithm}
\begin{algorithm}
\caption{CRC32 reverse algorithm}
\begin{algorithmic}
\REQUIRE{final CRC register value $r$,\ reversed data bits $a$}
\ENSURE{initial CRC register value $r^{'}$}
 \STATE $i \gets {a.length-1}$
 \STATE $crcreg \gets r$
 \WHILE{$i \geq 0$}
  \IF{$crcreg_{31} = 1$}
   \STATE $crcreg \gets {crcreg\ \oplus\ }$CRCPOLY
   \STATE RIGHTSHIFT($crcreg$)
   \STATE $crcreg_{0}= a_i \oplus 1$
  \ELSE
   \STATE RIGHTSHIFT($crcreg$)
   \STATE $crcreg_{0}= a_i$
  \ENDIF
  \STATE $i \gets {i-1}$
 \ENDWHILE
 \STATE $r' \gets crcreg$
\end{algorithmic}
\end{algorithm}

Assuming the initial CRC register value $r'$ and the data bits $a$ are available, the CRC algorithm can calculate the final value $r$.
If the final CRC register value $r$ and the calculated data $a$ are given, we can reverse the procedure of the CRC algorithm to obtain the initial value $r'$.
The CRC reverse algorithm \cite{b18} is given by Algorithm 2.

We can represent CRC algorithm and CRC reverse algorithm as functions where $r'$ and $r$ stand for the initial and final value of CRC register while $a$ stands for the calculated data bits.
If a set of CRC register values and data bits are given, the two algorithms are opposite computational processes, and the two functions hold simultaneously.

\begin{linenomath}
\begin{equation}
r=crc(r', a),\quad r'=crc\_reverse(r, a).
\end{equation}
\end{linenomath}

\subsubsection{How to reverse the unknown data from CRC32 value?}

The algorithms of CRC use forward or reverse methods to calculate the value of the CRC register from the data bits $a$.
Suppose we want to reverse the unknown data bits from the initial and final state values of CRC32.
Assuming the unknown data length $l$ is known, we can use brute-force search to find the possible original data from all $2^l$ data sequences.
When the data length $l$ is greater than 32 bits, the number of solutions is $2^{l - 32}$.
If the data length does not exceed 32 bits, the unknown data sequence is unique.
The relationship between the number of solutions $N_l$ and the data length $l$ can be expressed as follows:

\begin{align}
\begin{split}
N_l= \left \{
\begin{array}{ll}
    2^{l-32},             & l\ >\ 32\\
    1,                    & l \leq 32
\end{array}
\right.
\end{split}
\end{align}

To ensure the uniqueness of the results, we discuss the unknown data bits with a length of 32 bits.
In algorithms of CRC, data sequence $a$ is the independent variable of the functions.
We need to find the new connection between $a$ and CRC register value to calculate the unknown bits.
Assuming the length of data bits is the same as the length of CRC32, the CRC algorithm has properties \cite{b18} which can be written as:

\begin{linenomath}
\begin{equation}
crc(r_1,a_1) \oplus crc(r_2,a_2) = crc(r_1 \oplus r_2, a_1 \oplus a_2),
\end{equation}
\end{linenomath}

\begin{linenomath}
\begin{equation}
crc(r',a)=crc(a,r')=r.
\end{equation}
\end{linenomath}

\begin{figure}[H]
\subfigure[Forward calculation by CRC algorithm]{
\includegraphics[width=12cm]{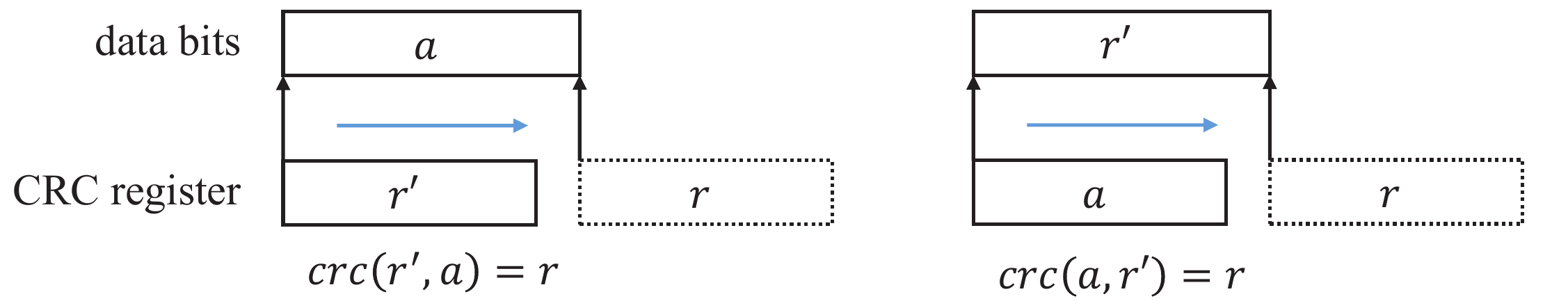}
}
\subfigure[Reverse calculation by CRC reverse algorithm]{
\includegraphics[width=12cm]{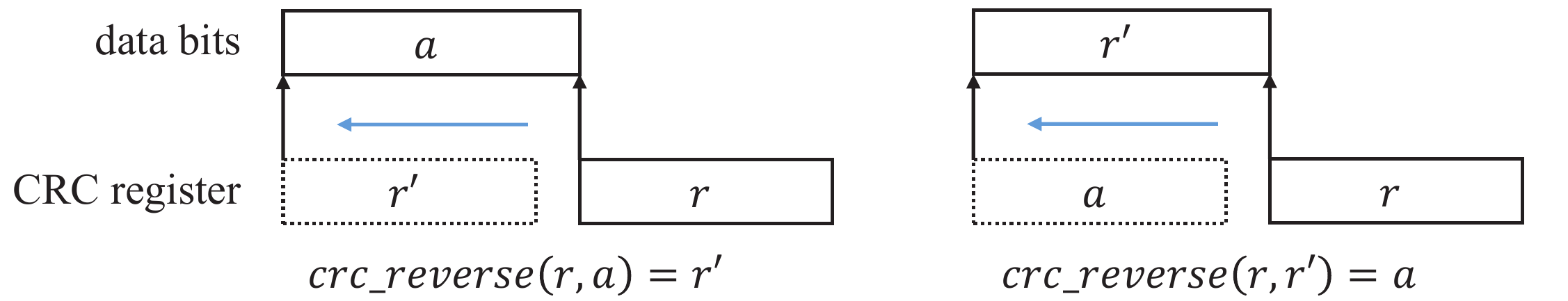}
}
\caption{Equation relations in CRC calculation. Because the CRC algorithm and the CRC reverse algorithm are different only in the calculation direction, the functions in (a) and (b) should hold simultaneously.\label{fig5}}
\end{figure}

Taking the equivalence relation of (6) and the correlation of algorithms (3), we can obtain a new equation (7) to calculate data bits $a$ exhibited in Figure~\ref{fig5}.

\begin{linenomath}
\begin{equation}
a = crc\_reverse(r,r').
\end{equation}
\end{linenomath}

% \begin{figure}[H]
% \subfigure[Forward calculation by CRC algorithm]{
% \includegraphics[width=10.5 cm]{newComparision1.pdf}
% }
% %\quad
% \subfigure[Reverse calculation by CRC reverse algorithm]{
% \includegraphics[width=10.5 cm]{newComparision2.pdf}
% }
% %\quad
% \caption{Equation relations in CRC calculation. Because the CRC algorithm and the CRC reverse algorithm are different only in the calculation direction, the functions in (a) and (b) should hold simultaneously.\label{fig5}}
% \end{figure}

\subsubsection{How to reverse the original data?}

We can divide the MAC frame into MAC Header $K$, CRC32 sequence $R$, and the unknown original data $a$.
Since the length of the original data is limited, theoretically, the original data can be obtained by brute-force search method.
The average number of enumerations is exponential to the original data length.
For the requirement of immediate communication and a high tag data transmission rate, brute-force search cannot satisfy them simultaneously.

If the length of the Frame Body field does not exceed 32 bits, we can get the original data by the method of reversing unknown data bits.
First, we use the algorithms of CRC to compute the initial and final values of CRC register $r'$ and $r$. 
According to the 802.11b protocol, the initial value of the CRC register for forwarding calculation is INITXOR while the final value of the CRC register $r'$ can be obtained from the FCS field.
Then, the original data can be calculated using the method of reversing unknown data bits.
The original packet can be obtained by replacing the backscatter data in the backscatter packet with the original data.
The procedure for reversing original data from the backscatter MAC frame is presented in Algorithm 3.

\begin{algorithm}
\caption{calculation in CRC reverse decoder}
\begin{algorithmic}
\REQUIRE{MAC Header $K$,\ CRC32 sequence $R$}
\ENSURE{original data $a$}
 \STATE \quad$r \gets R\ \oplus$ FINALXOR
 \STATE \quad$r' \gets crc($INITXOR$,K)$
 \STATE \quad$a \gets crc\_reverse(r,r')$ 
\end{algorithmic}
\end{algorithm}

CRC reverse decoder achieves efficient decoding computation in the case of transmission tag data length not exceeding 32 bits.
At the system level, the CRC reverse decoder is functionally identical to the receiver of a traditional backscatter system that receives the original packet.
In other words, the CRCScatter system reduces the hardware requirements by the CRC reverse decoding method.
At the principle level, the core for decoding is the presence of a bit sequence within the packet that constrains the transmitted data.
Because the CRC reverse decoder only uses the constraints of the FCS field, the maximum length of the tag data is 32 bits.
When the tag data length exceeds the limit, the reversed original data at the decoder will not satisfy the uniqueness of solutions.
The length of the tag data transmitted by the system can be extended when more constraints are applied to the packets.

\subsection{XOR and Differential decoder}
\begin{figure}[H]
\includegraphics[width=9 cm]{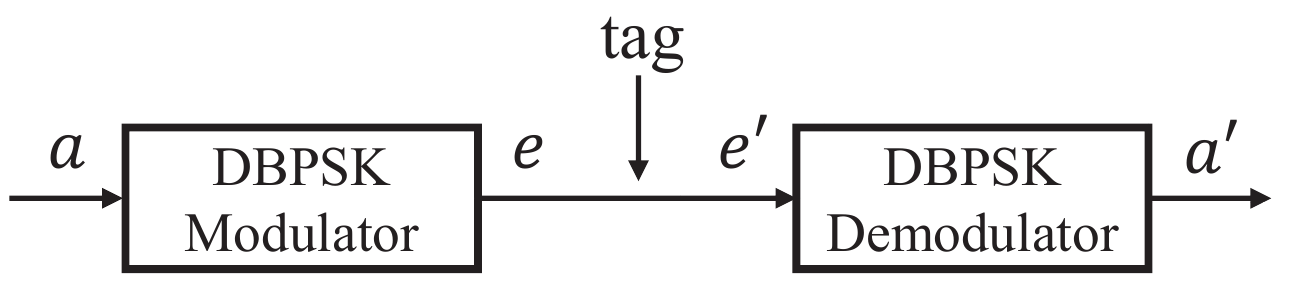}
\caption{The tag modulates the differential data rather than the original data, and tag data $t$ can be calculated by $e$ and $e'$.\label{fig6}}
\end{figure}

The 802.11 protocol specifies that the DSSS system uses the baseband modulation of DBPSK to provide the 1 Mbit/s data rate and solve the problem of phase ambiguity in BPSK.
In DBPSK modulation, the input original data $a$ is calculated by differential encoding to obtain differential data $e$, then the differential data $e$ is modulated by the conventional BPSK modulator.
In other words, we can assume that the 802.11b signal in our system transmits the differential data $e$ rather than the original data $a$.
Since the CRCScatter tag can modify the transmitted data by codeword translation, we set the differential data modified by the tag to $e'$ and the backscatter data to $a'$.
According to the decoding formula (1) and the process shown in Figure~\ref{fig6},  the tag data $t$ can be represented using differential data $e$ and $e'$.

\begin{linenomath}
\begin{equation}
t=e \oplus e',
\end{equation}
\end{linenomath}

In differential encoding, the transmitted data $a$ and the differential data $e$ should satisfy the encoding formula written as

\begin{linenomath}
\begin{equation}
e_i=e_{i-1}\oplus a_i.
\end{equation}
\end{linenomath}

We can use original data $a$ and backscatter data $a'$ to calculate tag data $t$ by differential encoding formula.
The tag data calculation needs to be discussed separately. The first tag data bit $t_0$ is calculated differently from the other tag data bits.
\begin{linenomath}
\begin{equation}
t_0 = e_0 \oplus e'_0 = a_0 \oplus a'_0,
\end{equation}
\end{linenomath}
\begin{linenomath}
\begin{equation}
t_i = e_i \oplus e'_i = (e_{i-1} \oplus a_i) \oplus (e'_{i-1} \oplus a'_i) = (a_i \oplus a'_i) \oplus (e_{i-1} \oplus e'_{i-1}) = (a_i \oplus a'_i) \oplus t_{i-1}.
\end{equation}
\end{linenomath}
%\addtolength{\topmargin}{0.011in}

Algorithm 4 presents the procedure to get correct tag data $t$ from backscatter data $a'$ and original data $a$ described in this section. 
\begin{algorithm}
\caption{calculation in XOR and Differential decoder}
\begin{algorithmic}
\REQUIRE{backscatter\ data $a'$,\ original\ data $a$}
\ENSURE{tag data $t$}
 \STATE $temp =a' \oplus a$
 \STATE $i \gets 0$
 \WHILE{$i\ <\ t.length$}
 \IF{$i = 0$}
   \STATE $t_i = temp_i$
  \ELSE
   \STATE $t_i = t_{i-1} \oplus temp_i$
  \ENDIF
  \STATE $i \gets i+1$
 \ENDWHILE
\end{algorithmic}
\end{algorithm}

In a real communication environment, bit errors in the received backscatter packet caused by noise will result in errors in decoding tag data.
Similar to the CRC32 sequence in the MAC frame, the tag can add redundancy check bits and piggyback them together with real tag data.
The CRCScatter decoder is required to check the tag data by redundant bits after tag data decoding.
In this way, the system can identify and discard incorrect results, and the accuracy of the decoding will be improved.

\section{Simulation results}

In this section, simulation results are presented to evaluate the performance of the proposed system. In our simulations, DBPSK modulation is used in 802.11b transmission and the length of the Frame Body field is set to 32 bits. The tag data length $N$ and the SNR will be varied to calibrate the results.

\begin{figure}[H]
\includegraphics[width=9 cm]{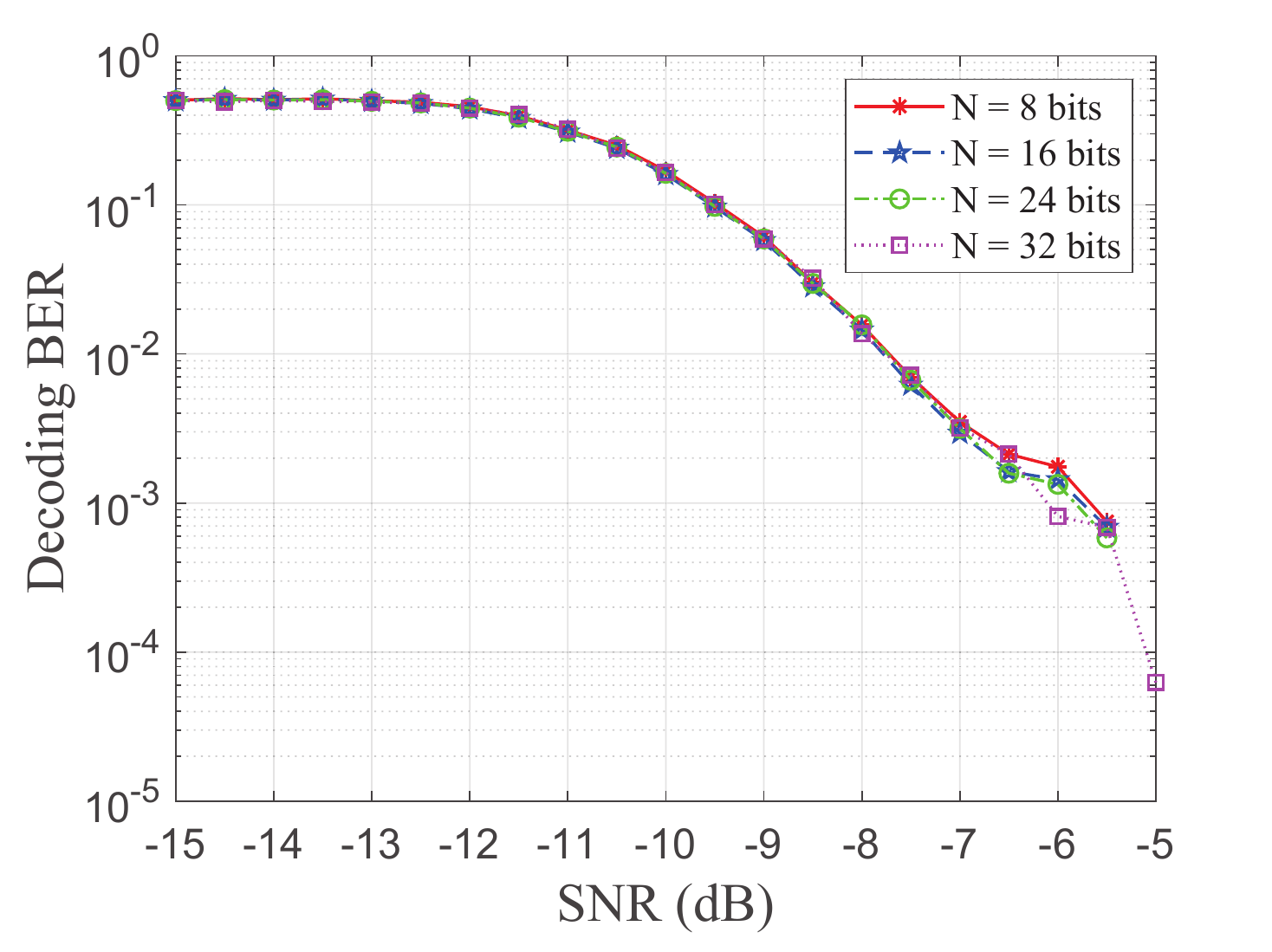}
\caption{The decoding BER versus SNR for different values of $N$. The decoding BER is independent of tag data length $N$.\label{fig7}}
\end{figure} 

First, we verify the effectiveness of the CRCScatter system. 
In the simulation, SNR can be set to different values through its relationship with $Eb/N0$.
Figure~\ref{fig7} shows the results of the decoding bit error rate of tag data versus SNR for different tag data lengths $N$. 
The SNR varies from $-15$ dB to $-5$ dB and the tag data length $N$ is fixed as 8, 16, 24, and 32 bits.
In Figure~\ref{fig7}, we can see that the decoding BER of tag data can be reduced by increasing the SNR. 
Nevertheless, decoding BER is similar in terms of the tag data length $N$.
From the figure, the system can achieve a BER level of $10^{-2}$ at $-7.5$ dB, which proves the effectiveness of the CRC reverse algorithm-based decoding method in the low SNR regime.
Moreover, the SNR does not affect the BER performance when the SNR is less than $-13$ dB. 
In this case, the system cannot decode correctly due to excessive noise interference.

\begin{table}[H] 
\caption{Comparison of two decoding methods.\label{tab2}}
\newcolumntype{C}{>{\centering\arraybackslash}X}
\begin{tabularx}{\textwidth}{CCC}
\toprule
\textbf{tag data length $N$ (bit)}	& \textbf{brute-force search $T_b$ (s)}	& \textbf{CRC reverse algorithm $T_c$ (s)}\\
\midrule
4		& 0.0026			& 0.0104\\
6		& 0.0049			& 0.0095\\
8		& 0.0194			& 0.0103\\
10		& 0.0586			& 0.0110\\
12		& 0.224			& 0.0118\\
14		& 0.896			& 0.0099\\
16		& 3.529			& 0.0111\\
18		& 38.707			& 0.0097\\
\bottomrule
\end{tabularx}
\end{table}
\unskip

Next, we test the decoding time of tag data using the brute-force search and CRC reverse algorithm.
Table~\ref{tab2} exhibits the results of average decoding time versus different tag data lengths for the algorithms.
We observe that the decoding time of brute-force search increases sharply when tag data length $N$ increases from 4 bits to 18 bits.
Due to excessive decoding time, the brute-force method is not able to meet the requirements of the real-time communication system.
Nevertheless, the tag data length $N$ does not affect the average decoding time of the CRC reverse algorithm, which is close to $1.0 \times 10^{-2}$ s.
Overall, the CRC reverse algorithm is superior to the brute-force search when the tag data length is long.
With the average decoding time of the CRC reverse algorithm and the maximum tag data length, we can estimate the maximum tag data rate of the system is 3.2 kbps, which is sufficient for intelligent meter reading systems, intelligent bracelets, and other micro IoT devices.

\begin{figure}[H]
\subfigure[The decoding BER versus SNR with N = 16 bits]{
\includegraphics[width=10cm]{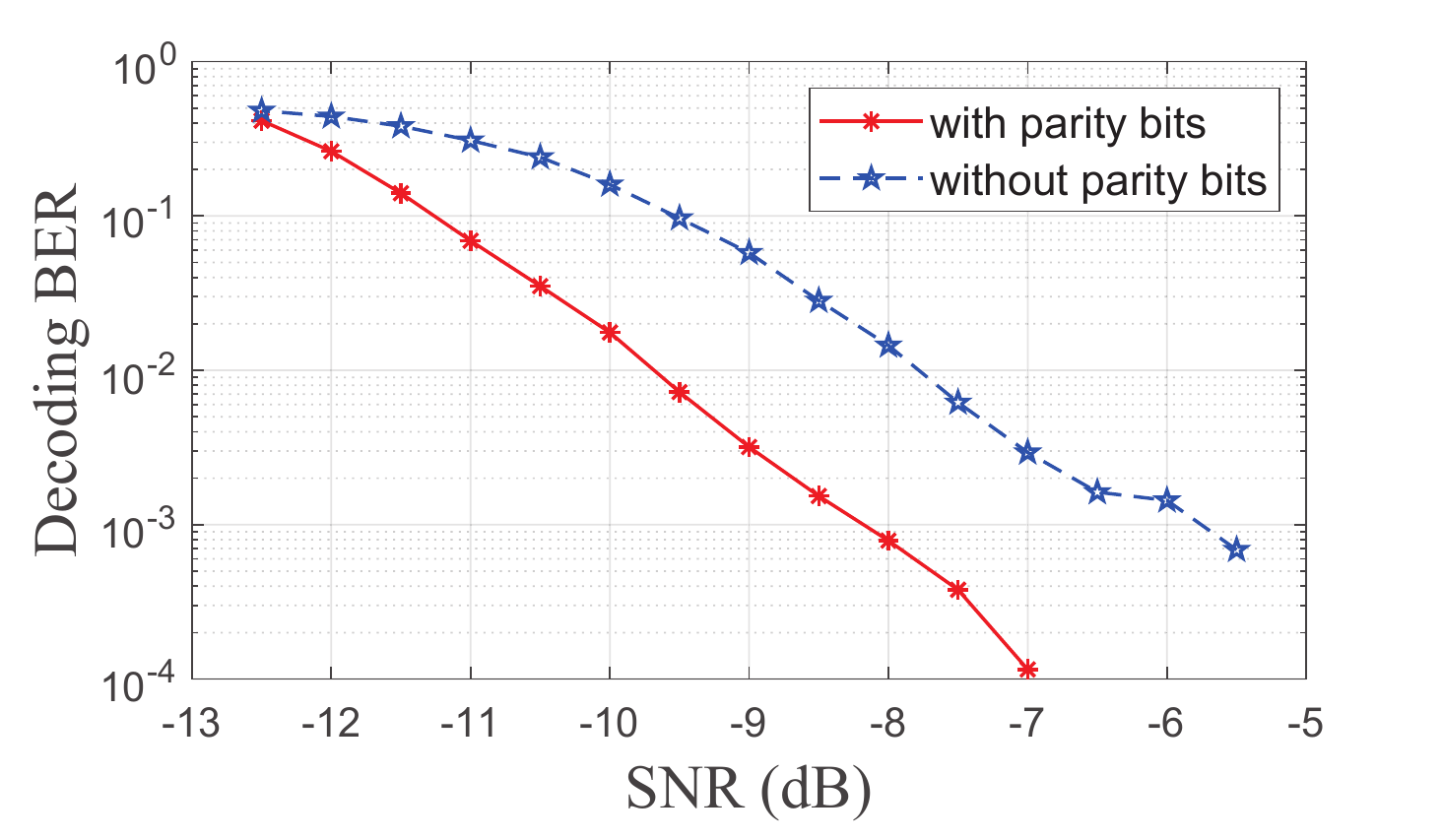}
}
\quad
\subfigure[The decoding BER versus N with SNR = -10 dB]{
\includegraphics[width=10cm]{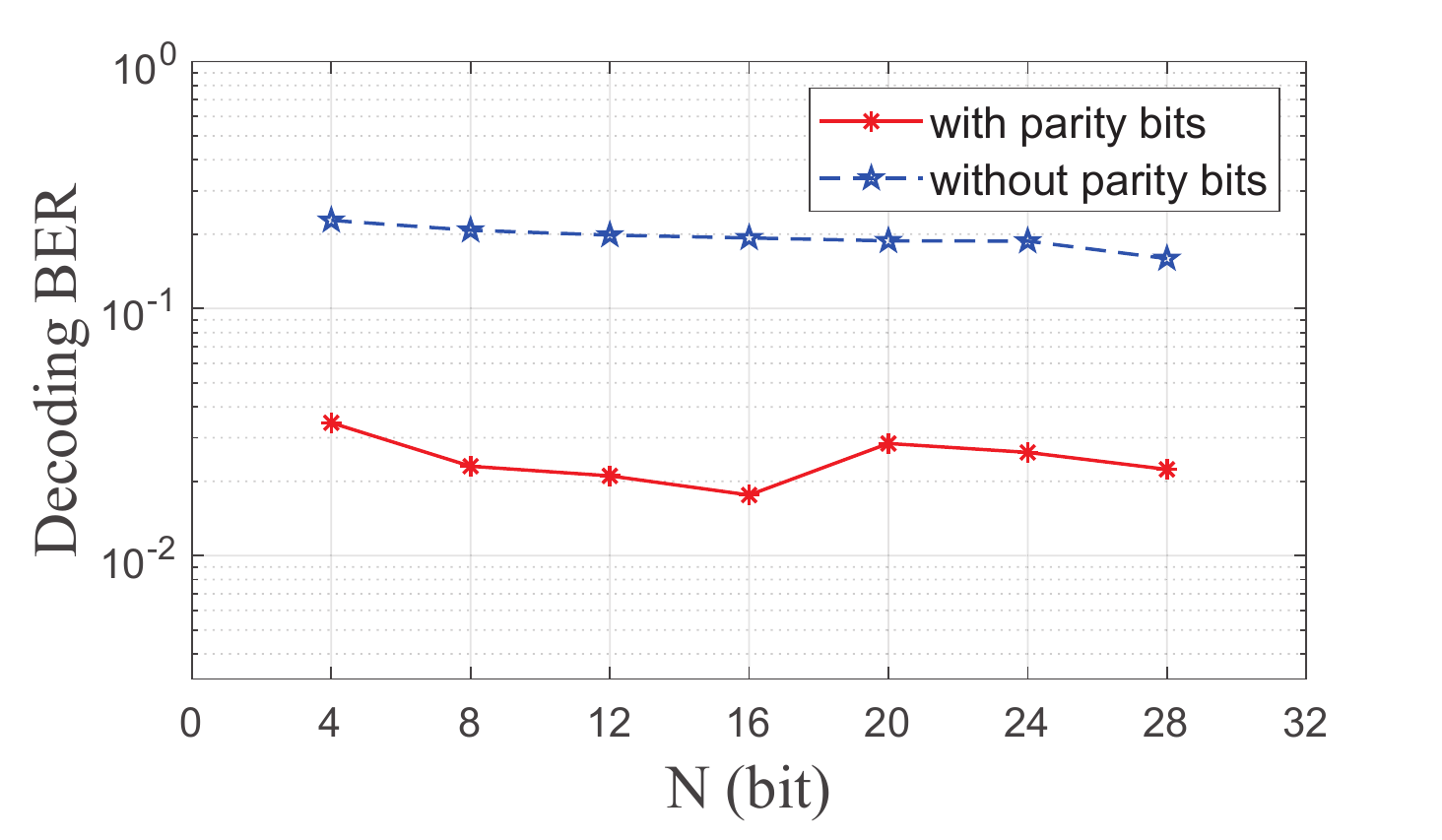}
}
\quad
\caption{Comparison of the decoding BER between tag data with four duplicate parity bits and tag data without check bits. Adding redundancy check bits can significantly reduce the decoding BER of the CRCScatter system.\label{fig8}}
\end{figure}

Finally, we test the performance of using redundant bits in tag data to reduce the decoding BER.
In the simulations, we use four duplicate parity bits as redundant bits for tag data. 
As shown in Figure~\ref{fig8} (a), the additional redundant bits can reduce the decoding BER when SNR is greater than $-12$ dB. 
In addition, when the SNR is higher than $-7$ dB, the improved method can achieve accurate decoding with tag data length $N = 16$ bits.
With the same SNR $=-10$ dB, Figure~\ref{fig8} (b) shows that the decoding BER of tag data with redundant bits is 9\% to 15\% to the BER without redundant bits.
We can conclude that the redundant bits can significantly reduce the system decoding error rate in the presence of noise interference.
Taking advantage of the stability of CRCScatter decoding time, adding redundant check bits will not affect the efficiency of system decoding.

\section{Discussion and conclusions}
In this paper, a novel backscatter communication system called CRCScatter is proposed to enable ambient WiFi backscatter communications with a single-AP receiver. CRCScatter requires neither an extra access point at the receiver, nor applying restrictions on the excitation source.
CRCScatter decoder achieves reversing the original excitation packet and decodes tag data by the received backscatter packet. The tag data can be decoded by following the procedure of CRC reverse, XOR decoding, and differential decoding. Simulation results validate the effectiveness of our proposed system. 

In future work, CRCScatter will be implemented and tested in real environments, and the decoding method will be extended to QPSK or other signals such as Bluetooth and ZigBee.

%%%%%%%%%%%%%%%%%%%%%%%%%%%%%%%%%%%%%%%%%%
\begin{adjustwidth}{-\extralength}{0cm}

\reftitle{References}

\end{adjustwidth}
\end{document}